\documentclass[sigconf]{acmart}
\AtBeginDocument{%
  }

\setcopyright{acmlicensed}
\copyrightyear{2026}
\acmYear{2026}
\acmDOI{XXXXXXX.XXXXXXX}
\acmConference[Conference'26]{Conference}{Date}{Place}

\usepackage{hyperref}

\hypersetup{
    pdfpagemode={UseOutlines},
    hypertexnames=false,
    colorlinks,
    urlcolor={blue},
    linkcolor={black},
    citecolor={black},
    pdfstartview={FitH},
    pdftitle={Prompt as a Data Type},
}
\usepackage{cleveref}

\usepackage{listings}
\usepackage{algorithm2e}

\usepackage{tikz}
\usepackage{pgfplots}
\pgfplotsset{compat=1.18}
\usetikzlibrary{arrows.meta,shapes,positioning,fit,backgrounds,calc,decorations.pathreplacing}

\usepackage{tcolorbox}
\tcbuselibrary{skins,breakable}
\usepackage{mdframed}

\usepackage{xparse}
\usepackage{xspace}
\usepackage{listings}
\usepackage{paralist}

\definecolor{navyblue}{RGB}{27,42,74}
\definecolor{acmblue}{RGB}{46,95,163}
\definecolor{lightblue}{RGB}{214,228,247}
\definecolor{codebg}{RGB}{245,247,251}
\definecolor{codered}{RGB}{180,0,0}
\definecolor{codegreen}{RGB}{0,120,0}
\definecolor{codegray}{RGB}{100,100,100}
\definecolor{alertbg}{RGB}{244, 244, 249}
\definecolor{alertborder}{RGB}{184, 219, 217}
\definecolor{defbg}{RGB}{240,246,255}
\definecolor{defborder}{RGB}{46,95,163}

\lstdefinelanguage{PromptQL}{
  morekeywords=[1]{SELECT,FROM,WHERE,CREATE,TABLE,TYPE,ALTER,SET,AS,WITH,ORDER,BY,IN,AND,OR,NOT,NULL,INTO,VALUES,INSERT,UPDATE,DELETE,DROP,INDEX,VIEW,MATERIALIZED,REFRESH,PROMPT,VERSION,TEMPLATE,INHERITS,ON,PRIMARY,KEY,FOREIGN,DEFAULT,RETURNS,FUNCTION,LANGUAGE,BEGIN,END,DECLARE,RETURN,IF,THEN,ELSE,ARRAY,TRIGGER,AFTER,FOR,EACH,ROW,EXECUTE,PROCEDURE},
  morekeywords=[2]{PROMPT,PromptSpec,ModelSpec,CachePolicy,STABLE,VOLATILE,IMMUTABLE,ASYNC,EXEC\_PROMPT,PROMPT\_MATCH,PROMPT\_DIFF,PROMPT\_AGGREGATE,SEMVER},
  morekeywords=[3]{INT,TEXT,FLOAT,BOOL,SERIAL,TIMESTAMP,INTERVAL,MAP,CHAR,VARCHAR,BIGINT},
  sensitive=true,
  morecomment=[l]{--},
  morecomment=[s]{/*}{*/},
  morestring=[b]',
  morestring=[b]"
}

\lstdefinestyle{sqlstyle}{
  basicstyle=\ttfamily\small,
  breaklines=true,
  columns=fullflexible,
  frame=single,
  language=SQL
}

\lstdefinestyle{textstyle}{
  basicstyle=\ttfamily\small,
  breaklines=true,
  columns=fullflexible,
  frame=single
}

\newtcolorbox{defbox}[1]{
  colback=defbg, colframe=defborder,
  fonttitle=\small\bfseries\color{navyblue},
  title={#1},
  left=4pt, right=4pt, top=3pt, bottom=3pt,
  arc=2pt, boxrule=0.8pt,
  before skip=6pt, after skip=6pt
}

\newtcolorbox{examplebox}[1]{
  colback=alertbg, colframe=alertborder,
  fonttitle=\small\bfseries\color{navyblue},
  title={#1},
  left=4pt, right=4pt, top=3pt, bottom=3pt,
  arc=2pt, boxrule=0.8pt,
  before skip=6pt, after skip=6pt
}

\newtcolorbox{findingbox}[1]{
  colback=lightblue!45, colframe=acmblue!45,
  fonttitle=\small\bfseries\color{navyblue},
  title={#1},
  left=4pt, right=4pt, top=3pt, bottom=3pt,
  arc=2pt, boxrule=0.7pt,
  before skip=5pt, after skip=5pt
}

\theoremstyle{definition}

\newcommand{\system}{PromptDB\xspace}
\newcommand{\prompttype}{\texttt{PROMPT}\xspace}
\newcommand{\evalop}{\texttt{EVAL}\xspace}

\newcommand{\optimizer}{PromptOpt\xspace}

\newcommand{\cmark}{\textcolor{codegreen}{$\checkmark$}}
\newcommand{\xmark}{\textcolor{codered}{$\times$}}

\newcommand{\eg}{\textit{e.g.},\ }
\newcommand{\ie}{\textit{i.e.},\ }

\begin{document}

\title[Prompt as a Data Type]{Prompt as a Data Type: In-Database LLM Prompt Management and Rewriting}

\author{Denis Mayr Lima Martins}
\authornote{Both authors contributed equally to this research.}
\affiliation{%
  \institution{Department of Computing and Mathematics\\University of Sao Paulo}
  \city{Ribeirão Preto}
  \state{São Paulo}
  \country{Brazil}
}\email{martins.denis@usp.br}
\orcid{0000-0002-8262-2369}

\author{Gottfried Vossen}
\affiliation{%
  \institution{Department of Information Systems\\University of M{\"u}nster}
  \city{M{\"u}nster}
  \country{Germany}}
\email{vossen@uni-muenster.de}

\renewcommand{\shortauthors}{Martins and Vossen}

\sloppy
\begin{abstract}
Large Language Models (LLMs) are increasingly used in database-backed applications to classify tuples, filter records using semantic predicates, extract structured attributes, and enrich query results. Yet the prompt that start these computations are typically stored outside the DBMS in  unstructured formats, making them invisible to query execution, metadata management, and optimization. Drawing on Stonebraker's \textit{QUEL as a Data Type} and the principles of reflective programming, this paper introduces \system, a database system that treats prompts as tuple-level database values. \system provides a logical \prompttype datatype whose values store a template, bindings to tuple attributes, model metadata, and task metadata. Relations may contain \prompttype attributes directly in base tables, or expose them through views over joined tuples. Users query prompt-valued attributes through generated evaluation views, while the system internally renders, rewrites, optimizes, and executes prompts through an \evalop operator. 
Making prompts database-visible creates a new optimization space. 
The key idea is to bring query-optimizer thinking to prompts: just as query optimizers exploit database metadata to rewrite SQL plans, \system exploits database metadata to rewrite prompts. 
We evaluate \system on synthetic and real-world data workloads across different tasks. The results show how database-guided rewriting improves output validity and yields favorable cost-quality trade-offs compared with static, manually written prompts. 
\end{abstract}

\maketitle

\section{Introduction}
\label{sec:intro}

Large Language Models (LLMs) are increasingly used as semantic operators over database tuples. A support system may classify a ticket from its text, a data-cleaning pipeline may normalize a messy value to a controlled domain, an analyst may filter orders using a natural-language predicate, and a procurement system may extract structured attributes from textual records. In all of these cases, the LLM computation is tuple-dependent: the prompt refers to database attributes, produces a value consumed by a query, and often must obey database-visible constraints such as valid labels or output formats. This paper promotes prompts to first-class citizens (i.e., data types) in a database.

\begin{figure*}[t]
    \centering
    \includegraphics[width=\linewidth]{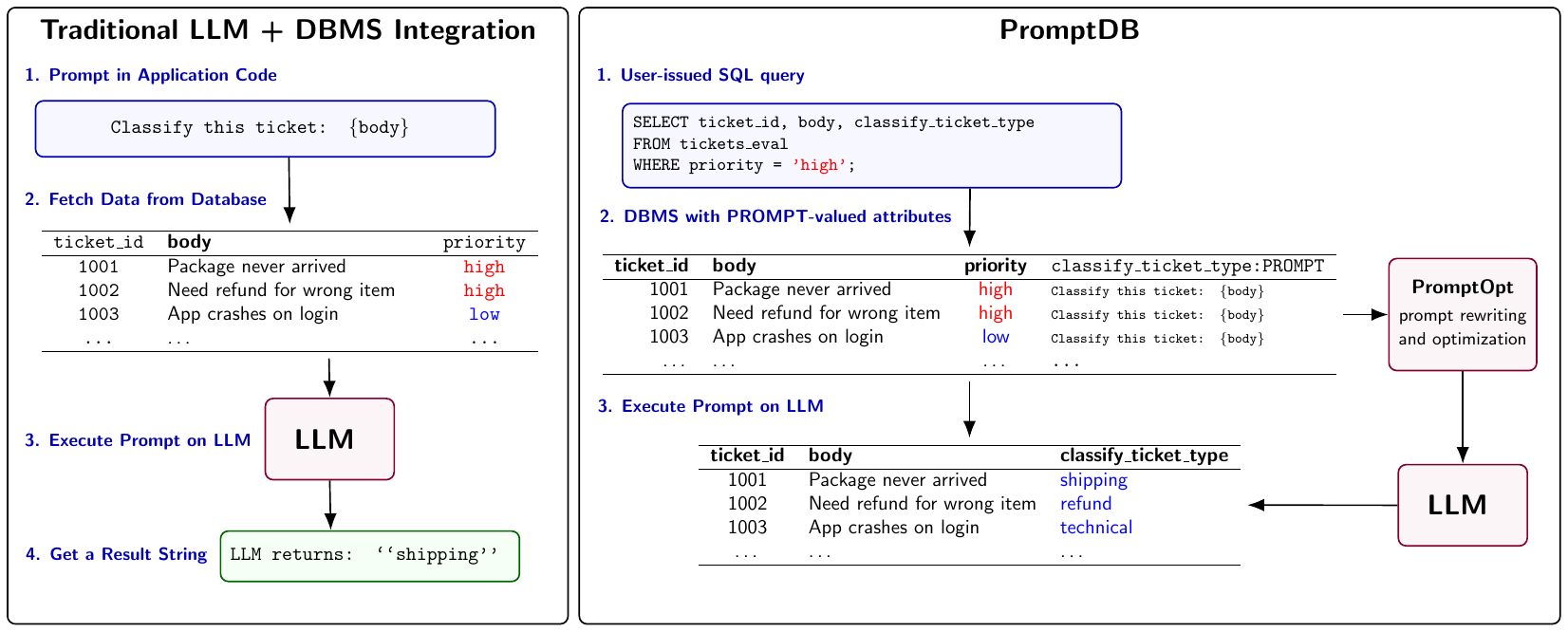}
    \caption{Motivating contrast between traditional prompt handling (left) and \system (right). A traditional system fetches tuples from the DBMS, invokes the LLM externally, and receives a result string outside the database. 
    In \system, the same prompts are represented as a tuple-level \texttt{PROMPT} attribute, here \texttt{classify\_ticket\_type} based on the tuple attribute \texttt{\{body\}}. 
    Users issue SQL to execute prompts, while \system manages prompt rewriting and optimization through \optimizer.}
    \label{fig:motivation}
\end{figure*}

\begin{examplebox}{Motivating Example}\label{ex:motivation}
\small
In a support-ticket database analysts use an LLM to classify customer complaints into one of four categories: \texttt{refund}, \texttt{delivery}, \texttt{technical}, or \texttt{other}. Today, the prompt might be constructed in application code as follows:

\begin{lstlisting}[numbers=none]
prompt = f"""
Classify this customer support ticket: {message}
"""
response = openai.Completion.create( engine="gpt-4", prompt=prompt)
\end{lstlisting}

Although the database already stores the valid category domain, the prompt does not expose it to the model. As a result, the LLM may return outputs such as ``The customer wants their money back,'' which may be semantically reasonable but are not valid database values.
\end{examplebox}

Despite this, the main artifact that controls LLM behavior, the prompts, are usually managed outside the DBMS. They appear as notebook cells, prompt templates inside application code, or workflow configuration files.  As a result, the database system cannot inspect, rewrite, explain, or optimize them. Existing SQL UDF-based approaches can call an LLM from SQL, but the prompt remains an opaque argument to a function. The DBMS sees the function call, not the prompt as a manageable value.

Inspired by Stonebraker's \textit{QUEL as a Data Type}~\cite{DBLP:conf/sigmod/StonebrakerAHR84} and the notion of treating query-language commands as data~\cite{10.1145/153850.153852}, we argue that prompts should be treated as first-class data types. We define a model where prompts are stored as \emph{intensional values} in table columns (\ie a custom data type called \prompttype). We implement this idea in \system, a database-guided prompt rewriting system for prompt-valued attributes. The key contribution is to bring query-optimizer thinking to prompts. A relational optimizer uses schema, constraints, statistics, and cost estimates to choose better query plans. \system applies the same principle to prompt-valued attributes: database context is used to produce prompts that are more valid and robust. The overall idea is depicted in \Cref{fig:motivation}. 

While prior work has focused on LLM-as-a-service or application-level prompt management, \textbf{\system is a database-centric approach that treats prompts as structured, queryable entities}. In essence, this offers a unified interface for SQL users to call and optimize LLM prompts without leaving the database, which opens new avenues for research in AI-native databases and prompt engineering at scale. Table~\ref{tab:abstractions} summarizes why existing abstractions are insufficient.

\begin{table}[t]
\centering
\caption{Comparison of prompt-related abstractions.}
\label{tab:abstractions}
\resizebox{\linewidth}{!}{%
\begin{tabular}{lcccc}
\toprule
Abstraction & Typed & Optimizable & Versioned & Rewrite \\
\midrule
Text/JSON type & \xmark & \xmark & Limited & \xmark \\
Application code & \xmark & No & External & \xmark \\
UDF & Weak & Limited & Manual & Limited \\
Semantic operator & \cmark & Limited & Limited & Limited \\
\prompttype (ours) & \cmark & \cmark & \cmark & \cmark \\
\bottomrule
\end{tabular}
}
\end{table}


This paper makes the following contributions:

\begin{enumerate}
    \item We introduce a \prompttype data type for executable prompt-valued attributes.
    \item We present a formal model of prompt-valued relations, prompt evaluation, database-guided prompt rewriting, and cost-aware prompt optimization.
    \item We implement \system on top of DuckDB using \prompttype attributes, generated evaluation views, an internal \evalop operator and \optimizer.
    \item We evaluate \system on synthetic and real-world data showing task-quality improvements, and prompt cost-quality trade-offs.
\end{enumerate}

The rest of this paper is organized as follows: \Cref{sec:formalmodel} introduces the formal model. \Cref{sec:systemdesign} presents the design of \system; \Cref{sec:experiments} evaluates our approach through experiments on two datasets; \Cref{sec:related} discusses related work; and \Cref{sec:conclusion} concludes with future directions.

\section{Formal Model}
\label{sec:formalmodel}

This section formalizes \system as a data model and execution model for prompt-valued relations.

\paragraph{\bf Prompt-Valued Relations}

Let a database contain a set of relations $\mathcal{D} = \{R_1,\ldots,R_n\}$.
Each relation $R$ has a schema $\mathsf{sch}(R) = A_1:\tau_1,\ldots,A_m:\tau_m$. 
In a conventional database, each $\tau_i$ is a scalar, structured, or collection type. \system extends the type system with a logical prompt, named $\prompttype$.

A relation may therefore contain ordinary attributes and prompt-valued attributes:
\begin{equation*}
    \mathsf{sch}(R) = A_1:\tau_1,\ldots,A_k:\tau_k, P_1:\prompttype,\ldots,P_\ell:\prompttype.
\end{equation*}

A tuple $t \in R$ contains ordinary values $t[A_i]$ and prompt values $t[P_j]$. Prompt attributes may appear in base tables, views, or materialized query results. This modeling is important for relational workloads, since a prompt may naturally belong not to a base tuple, but to a derived tuple produced by a join or projection.

\paragraph{\bf Prompt Values}

A \prompttype value is a self-describing executable object: 
\begin{equation}
    P = \langle T, B, M, D \rangle,
\end{equation}

\noindent where $T$ is a prompt template containing variables, $B$ maps template variables to tuple attributes, $M$ specifies the model or model family, $D$ is the output domain or output specification. 

For example, a support-ticket tuple may contain:

\begin{lstlisting}[numbers=none]
template: "Classify this ticket: {{body}}"
bindings: { body -> body }
model: "llama-3.2-3b-instruct"
output_domain: ["delivery", "refund", "technical", "other"]
\end{lstlisting}


\paragraph{\bf Tuple Rendering}

Let $t$ be a tuple and $P = \langle T,B,M,D,\rangle$ a prompt value. Rendering substitutes template variables using the binding map:
\begin{equation}
    \mathsf{render}(P,t) =
T[B(v_1) \mapsto t[B(v_1)],\ldots,B(v_r) \mapsto t[B(v_r)]].
\end{equation}

For example:

\begin{lstlisting}[numbers=none]
Template:
Classify this ticket: {{body}}
Tuple:
body = "My package never arrived."
Rendered prompt:
Classify this ticket: My package never arrived.
\end{lstlisting}

Rendering is intentionally defined separately from rewriting. \system may first transform the template and only then render it against the tuple.

\paragraph{\bf Prompt Rewriting Context}

Prompt execution is parameterized by a database-derived context:

\begin{equation}
\Omega = \langle S, C, Q, T_s \rangle.
\end{equation}

Here:

\begin{itemize}
    \item $S$ is schema context, including column names, types, comments, and optional profiled descriptions;
    \item $C$ is constraint context, including output domains and validity requirements;
    \item $Q$ is query context, including projected attributes, predicates, and task bindings;
    \item $T_s$ is statistics context, including sample values, label distributions, and representative examples.
\end{itemize}

The context is stored as JSON next to each prompt attribute in the prototype's generated relations and views.

\paragraph{\bf Prompt Evaluation}

Conceptually, evaluating a prompt attribute returns a scalar SQL value, i.e., $\mathsf{EVAL}(P,t,\Omega) \rightarrow y$, where $y$ is a string, label, Boolean-like value, or structured textual value depending on the task. More explicitly:
\begin{equation}
\mathsf{EVAL}(P,t,\Omega) =
\mathsf{LLM}
\left(
\mathsf{render}
\left(
\rho_{\sigma}(P,\Omega),t
\right)
\right),
\end{equation}

\noindent where $\rho_{\sigma}$ is the rewrite strategy selected by the prompt value's default strategy $\sigma$. If $\sigma=$ \texttt{PromptOpt}, the strategy is selected by \optimizer.

{\bf Prompt rewriting} transforms a prompt value $P$, given a rewriting function $\rho_r$, into an execution prompt $\rho_r(P,\Omega) = P'$, where $P'$ typically has the same bindings and model metadata as $P$, but a modified template $T'$. A rewrite may inject constraints, reduce tuple fields, add examples, or impose an output format.

A set of rewrite rules defines a rewrite plan:
$\pi = [r_1,\ldots,r_k]$. 
Applying a plan produces $P_\pi = \rho_{r_k}(\cdots \rho_{r_1}(P,\Omega)\cdots)$.

\paragraph{\bf Prompt Optimization}

Different rewrite plans have different quality and cost behavior. Adding few-shot examples may improve quality but increase tokens. Projecting columns may reduce cost but risk removing useful context. Therefore, \system treats prompt rewrites as alternative execution plans.

Let $\mathcal{R}(P)$ be a finite set of candidate rewritten prompts. \optimizer selects:
\begin{equation}
P^* =
\arg\max_{P_i \in \mathcal{R}(P)}
\left(
\hat{Q}(P_i,\Omega)
-
\lambda \hat{C}(P_i,\Omega)
\right),
\end{equation}
\noindent where $\hat{Q}$ is an estimated quality score, $\hat{C}$ is an estimated cost, and $\lambda$ controls the quality-cost trade-off.

In our implemented prototype, $\hat{C}$ is estimated from rendered prompt length and expected output length. $\hat{Q}$ is a lightweight heuristic based on rule applicability. For example, constraint injection is considered useful when the output domain is known, projection is useful for wide relational tuples, and few-shot examples are useful when examples are available. Learned or calibration-based quality estimation could be applied here.

\section{System Implementation}
\label{sec:systemdesign}

We implemented \system on top of DuckDB. The prototype is designed to demonstrate the data model and optimization opportunities while avoiding invasive modifications to DuckDB internals.

\paragraph{\bf Logical \prompttype Datatype}
The prototype defines \prompttype as a DuckDB structured type:

\begin{lstlisting}[language=SQL,numbers=none]
CREATE TYPE PROMPT AS STRUCT(
    template VARCHAR,
    bindings MAP(VARCHAR, VARCHAR),
    model VARCHAR,
    task VARCHAR,
    output_domain VARCHAR[],
    name VARCHAR,
    default_strategy VARCHAR
);
\end{lstlisting}

Each \prompttype value is self-describing. To facilitate the implementation, we store the template, tuple bindings, model name, task, output domain, prompt name, and default rewrite strategy directly as part of \prompttype.

\paragraph{\bf Prompt Attributes in Tables and Views}
For single-table workloads, \system stores \prompttype attributes directly in the base relation. For example, the synthetic support-ticket relation contains:

\begin{lstlisting}[language=SQL,numbers=none]
tickets(
  ticket_id BIGINT,
  body VARCHAR,
  priority VARCHAR,
  declared_category VARCHAR,
  semantic_value_normalization_prompt PROMPT,
  semantic_filtering_prompt PROMPT,
  attribute_extraction_prompt PROMPT
)
\end{lstlisting}

\paragraph{\bf Generated Evaluation Views}

DuckDB does not automatically execute a UDF when a structured column is projected. Therefore, the prototype uses generated evaluation views. For every relation with \prompttype attributes, \system creates a corresponding \texttt{\_eval} view. The view exposes prompt columns under the same names, but internally expands them into \evalop calls. For example, \texttt{tickets\_eval} contains expressions of the form:

\begin{lstlisting}[language=SQL,numbers=none]
EVAL(semantic_filtering_prompt,
  semantic_filtering_row_json,
  semantic_filtering_context_json,
  struct_extract(semantic_filtering_prompt, 'default_strategy'),
  '[]'
) AS semantic_filtering_prompt
\end{lstlisting}

\newpage
The user can then write:

\begin{lstlisting}[language=SQL,numbers=none]
SELECT ticket_id, body, semantic_filtering_prompt
FROM tickets_eval
WHERE semantic_filtering_prompt = 'yes';
\end{lstlisting}

This hides \evalop from the user while preserving an explicit execution operator inside DuckDB.

\paragraph{ \bf Internal \evalop Operator}

The \evalop operator is registered as a DuckDB scalar UDF. Its logical signature is:

\begin{lstlisting}[numbers=none]
EVAL(prompt PROMPT,
     row_json VARCHAR,
     context_json VARCHAR,
     strategy VARCHAR,
     examples_json VARCHAR) -> VARCHAR
\end{lstlisting}

At runtime, \evalop performs the following steps: 
\begin{inparaenum}
    \item deserialize the \prompttype value;
    \item deserialize the tuple context and optimization context;
    \item select a rewrite strategy;
    \item apply \optimizer;
    \item render the final prompt against the tuple;
    \item call LM Studio through a local client;
    \item return the model output as a SQL string;
    \item append execution metadata to an in-memory log.
\end{inparaenum}

The execution metadata includes the requested strategy, actual strategy, applied rules, estimated cost, rendered prompt, prediction, latency, input tokens, and output tokens.

\paragraph{\bf Rule-based Prompt Rewriting}

Similar to query optimization approaches, \system employs a rule-based prompt rewriting.




\paragraph{$R_{CI}$: Constraint Injection.}
Inspired by \cite{geng2025jsonschemabench}, this rule injects valid output values derived from database constraints to make the expected output domain explicit to the LLM. 

\begin{lstlisting}[numbers=none]
Original:
Classify this ticket: {{body}}

Rewritten:
Classify this ticket.
Return exactly one of: refund, delivery, other.
\end{lstlisting}

\paragraph{$R_{QCP}$: Query-Aware Column Projection}
This rule adapts projection pushdown to prompt execution: only tuple attributes relevant to the prompt task are rendered. This mirrors classical relational optimization and recent prompt-compression/context-pruning work~\cite{fang2025attentionrag}, which show that reducing irrelevant context can lower cost and sometimes improve quality.

\begin{lstlisting}[numbers=none] Original: 
Tuple: orderkey=1024, custkey=91, orderpriority=1-URGENT, shipmode=AIR, commitdate=1995-03-12, receiptdate=1995-03-20 

Rewritten: 
Relevant: orderpriority=1-URGENT, shipmode=AIR, commitdate=1995-03-12, receiptdate=1995-03-20 
\end{lstlisting}

\paragraph{$R_{OFM}$: Output Format Minimization.}
This rule enforces concise, parseable outputs. Recent work shows that downstream systems benefit when LLM outputs are concise, parseable, and aligned with the expected value~\cite{geng2025jsonschemabench}.

\begin{lstlisting}[numbers=none] Original: 
Classify: {{body}} 

Rewritten: 
Classify: {{body}} 
Return label only. 
\end{lstlisting}

\paragraph{$R_{FSE}$: DB-Selected Few-Shot Examples}
This rule instantiates in-context learning inside the DBMS: examples are selected from database tuples and inserted into the prompt as demonstrations. This follows prior work~\cite{liu2021makesgoodincontextexamples,luo2024incontextlearningretrieveddemonstrations} showing that few-shot prompting is effective and that the choice of demonstrations strongly affects performance.

\begin{lstlisting}[numbers=none] Original: Is delayed? {{tuple}} 

Rewritten: 
Examples: urgent, late -> yes low, on-time -> no 
Is delayed? {{tuple}} 
\end{lstlisting}

The implementation treats a rewrite plan as an ordered list of rules. This way, \system does not treat prompt engineering as ad hoc application logic. Instead, it recasts recurring prompt-engineering patterns as database rewrite rules over typed prompt values.


For efficiency, \optimizer considers a fixed set of rewrite candidates:
\begin{description}
    \item[NoRewrite:] No rules are applied.
    \item[Constraint+Format:] Rules $R_{CI}$ and $R_{OFM}$ are applied.
    \item[Projection:] Rule $R_{QACP}$ is applied.
    \item[FewShot:] Rules $R_{CI}$, $R_{OFM}$, and $R_{FSE}$ are applied.
    \item[AllRules:] $R_{CI}$, $R_{QACP}$, $R_{OFM}$, and $R_{FSE}$ are applied.
\end{description}

For each candidate, \optimizer estimates quality and cost, then selects the best candidate according to $\hat{Q} - \lambda \hat{C}$. Note that the optimizer is intentionally lightweight.

\section{Experiments and Results}\label{sec:experiments} 


We implement \system as a modular Python prototype with \texttt{DuckDB~1.5.4}, operating in fully in-memory mode (\texttt{":memory:"}) as the default backend. The backend interface is replaceable, allowing future implementations over SQLite or PostgreSQL. 
Prompt execution uses LMStudio to call a lightweight \texttt{llama-3.2-3b-instruct}. The LLM backend is also replaceble (\eg allowing for ollama or OpenAI API calls). All inference calls use \texttt{temperature} set to 0.0 for determinism; \texttt{max\_tokens} is set to 128. All experiments run on a single commodity workstation. 

\paragraph{Goal of the experiments.} We therefore focus on three questions: 
\begin{inparaenum}
     \item whether database-guided prompt rewriting improves end-to-end task quality,
     \item which rewrite rules contribute to the observed behavior, and 
     \item whether \optimizer can reduce prompt cost while retaining competitive task quality.
\end{inparaenum}

\paragraph{Datasets.}

 We use three datasets representing different database settings.

 \begin{description}
 \item[Synthetic tickets:] A synthetic support-ticket dataset with controlled labels, entities, and references. 
\item[Car Evaluation:] Available at OpenML, this dataset comprises a single database table including attributes such as buying price, maintenance cost, passenger capacity, luggage size, safety, and acceptability. It is useful for controlled-domain semantic normalization and filtering.
\item[TPC-H:] We used DuckDB's TPC-H generator. Prompt-bearing tuples are derived from relational joins over orders and lineitems. TPC-H is used to demonstrate prompt-valued attributes over multi-table relational tuples.
 \end{description}

Each dataset is evaluated across multiple prompt tasks.

\paragraph{Tasks.}
We define three distinct prompt-based tasks. First, \textbf{Attribute Extraction} requires the model to map a given data tuple to a controlled vocabulary label, testing its capacity for constrained classification. Second, \textbf{Semantic Filtering} assesses whether the model can determine if a tuple satisfies an arbitrary natural-language condition; in this case, the ground truth is derived deterministically from simple predicate logic or predefined categorical attribute values. Finally, \textbf{Semantic Value Normalization} challenges the model to extract and structure specific attributes from either a raw data tuple or its rendered textual representation, ensuring that the extracted output aligns precisely with defined database schema attributes. For each task, ground truth is deterministic, derived from simple predicates or selected attribute values.



\paragraph{Metrics.} For semantic value normalization and semantic filtering, we report accuracy, macro F1, and valid-output rate. For attribute extraction, we report attribute-level precision, recall, F1, and exact-match rate. For optimization experiments, we also report input tokens, output tokens, latency, and cost-adjusted quality.

\paragraph{Baselines and Strategies} 
We benchmark performance against the three prompt execution strategies. The simplest baseline is the \textbf{Static} strategy, which utilizes the unmodified initial prompt provided by the developer, serving as a foundational measure of raw model capability. We further introduce the \textbf{PromptRewrite} strategy, which employs a fixed, database-guided rewrite mechanism. Finally, we apply \textbf{\optimizer} cost-aware strategy.


\subsection{End-to-End Task Quality} 

\begin{figure}[t]
    \centering
    \includegraphics[width=0.95\linewidth]{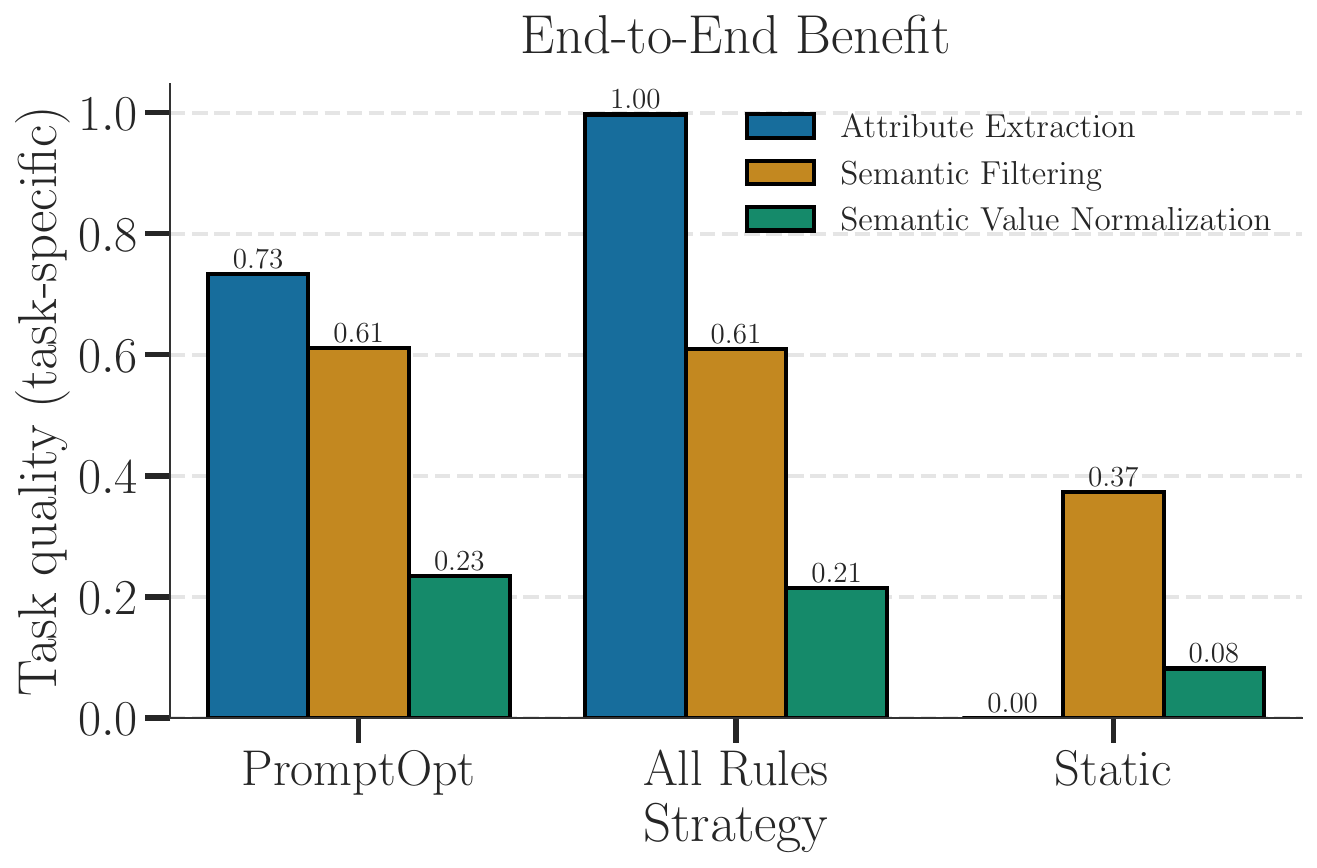}
    \caption{End-to-end task quality. 
    Applying all rules yields the strongest result, while \optimizer achieves competitive performance with a cheaper selected plan.}
    \label{fig:end_to_end_benefit}
\end{figure}

Figure~\ref{fig:end_to_end_benefit} compares three execution strategies: static prompting, fixed application of all rewrite rules, and \optimizer. 
Static prompting represents the application-level baseline in which the prompt is rendered and executed without database-guided rewriting. 
The all-rule strategy applies the full rule set. 
\optimizer selects a rewrite strategy using its cost-quality objective.

Overall, obtained results shows that prompt-valued attributes enable meaningful database-guided execution. The gains are strongest when the task has clear output-structure requirements, as in attribute extraction and semantic filtering. In contrast, the weaker results for semantic value normalization show that prompt optimization must be task-aware, rather than relying on a single fixed rule set for all prompt-valued attributes.

\subsection{Rule Ablation} 

\begin{figure}[t]
    \centering
    \includegraphics[width=0.95\linewidth]{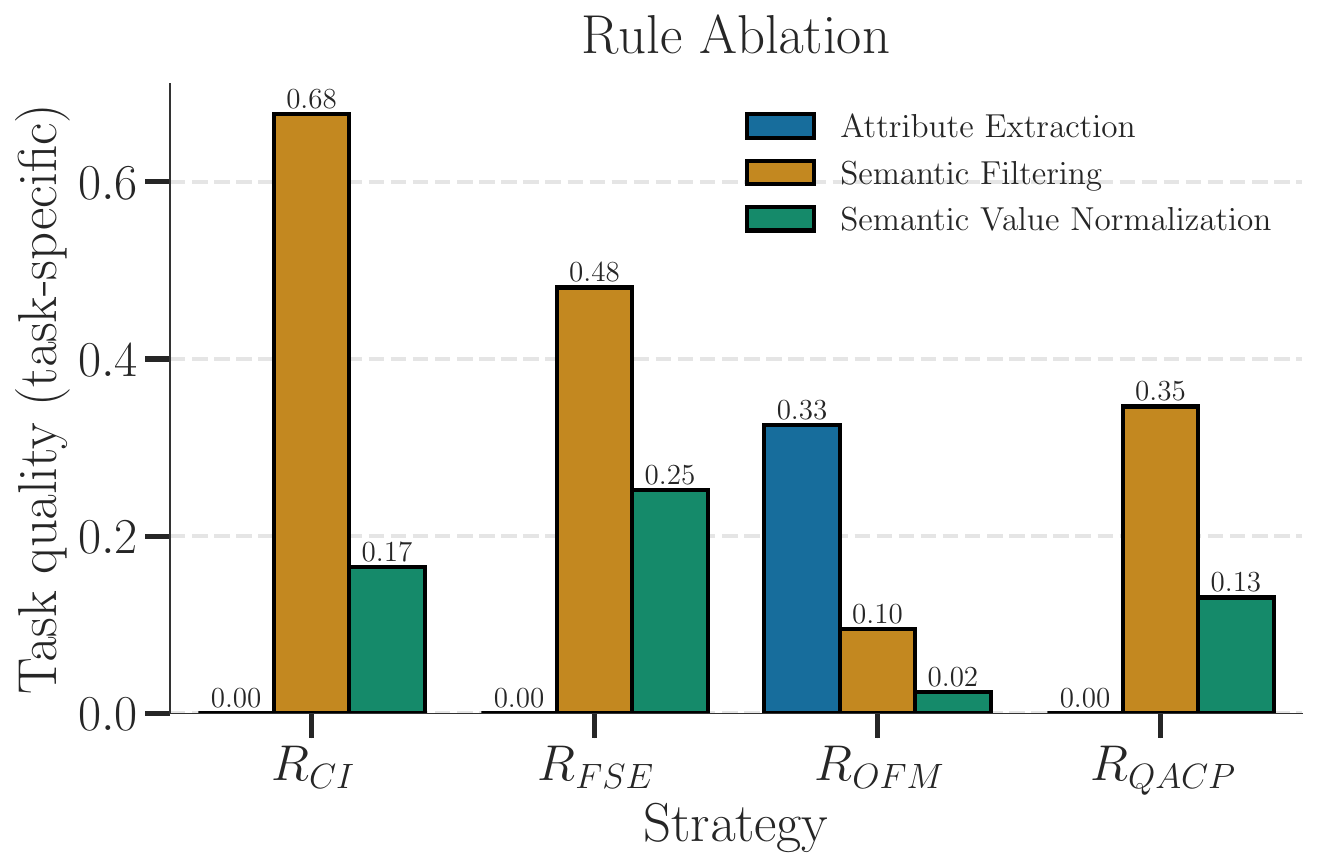}
    \caption{Rule ablation. The results show that rewrite rules are task-sensitive and that individual rules often address different failure modes.}
    \label{fig:rule_ablation}
\end{figure}

Figure~\ref{fig:rule_ablation} isolates the effect of individual rewrite rules. The purpose of this experiment is to understand which failure modes are addressed by each rule. 
The ablation results show that no single rewrite rule dominates across all tasks. Instead, each rule helps under different task conditions. This motivates \optimizer and future learned variants that choose rewrite plans based on task-specific utility rather than applying a fixed strategy.

\subsection{Cost-Quality Trade-Off} 

\begin{figure}[t]
    \centering
    \includegraphics[width=0.99\linewidth]{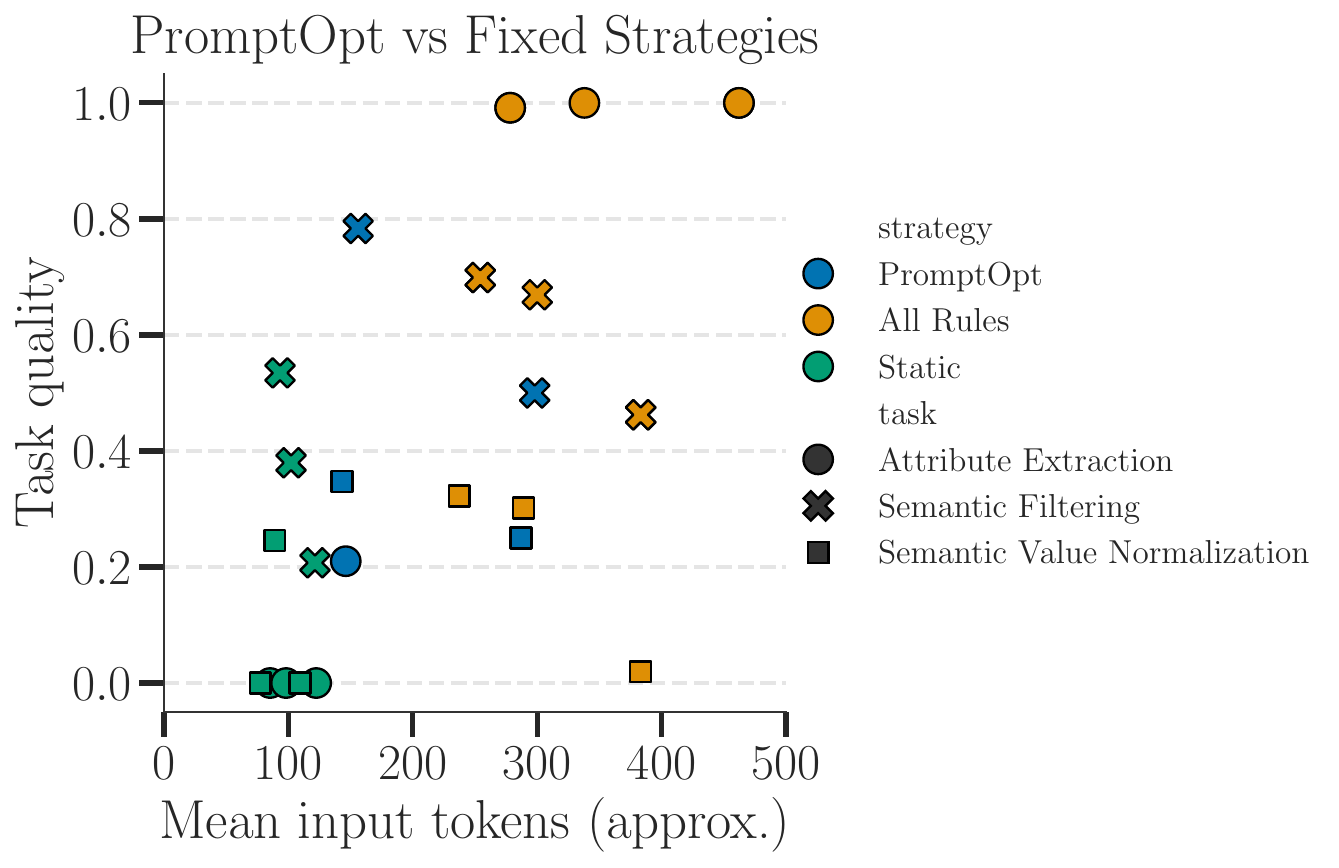}
    \caption{Cost-quality trade-off between \optimizer and fixed prompting strategies. 
    \optimizer occupies an intermediate region by selecting lower-cost rewrite plans while retaining competitive quality for some tasks.}
    \label{fig:promptopt}
\end{figure}

Figure~\ref{fig:promptopt} compares \optimizer with fixed strategies in terms of task quality and approximate mean input tokens. 
Rather than always applying the most expensive prompt plan, \optimizer attempts to select a plan whose expected quality justifies its token cost. 
However, for attribute extraction, \optimizer does not always choose sufficiently rich rewrite plans. 
This suggests that the current heuristic quality estimator should be made task-aware. 
In particular, structured-output tasks should assign higher utility to plans that include expected-key constraints, strict output formatting, and format-perfect examples.

\subsection{Discussion}
\label{subsec:expdiscussion}

Database-guided prompt rewriting improves over static prompting for tasks that require constrained or structured outputs.  This supports the core design decision of \system, in which prompts should be represented as database-visible values rather than opaque application strings.

However, rewrite effectiveness is task-dependent. \optimizer demonstrates the feasibility of cost-aware prompt-plan selection, but its current heuristic estimator is incomplete. These results motivate future work on task-aware, calibrated, or learned models.

Once prompts are stored as tuple-level database values, the DBMS can inspect their metadata, rewrite them using database context, and optimize their execution under quality-cost trade-offs. 
The current prototype exposes this optimization space and demonstrates its potential, while also identifying concrete directions for improving the optimizer and rewrite rules.



\section{Related Work}
\label{sec:related}

Frameworks such as DSPy~\cite{khattab2024dspy} and other prompt optimization approaches \cite{ding2022openprompt,bach2022promptsource} automate or assist prompt engineering. These approaches treat prompts as unstructured assets, lacking the queryability of \system. Furthermore, \system differs by treating database metadata as the primary source of prompt-rewrite signals.

Recent work explores LLM operators within query pipelines.
\texttt{EVAPORATE}~\cite{arora2023llm} generates extraction functions from natural
language; it treats prompts as ephemeral runtime strings with no
persistence.  LOTUS~\cite{patel2024semantic} introduces semantic operators
(\texttt{sem\_filter}, \texttt{sem\_join}) as DataFrame-level
abstractions, where prompts are operator parameters, not database values. Closest to our work, \texttt{SPEAR}~\cite{ugur2026firstclass} proposes structured prompt views, prompt algebra, adaptive prompt refinement, and policy-driven control. Its emphasis is on treating prompts as structured, composable, and refinable pipeline objects. \system is complementary: it focuses specifically on how database metadata can rewrite prompt templates for prompt-valued attributes. Unlike SPEAR, \system studies a database-centric question: what can schema, constraints, statistics, and feedback do for prompt rewriting?
Moreover, \system rewrites prompt templates directly using database-native context.

\section{Conclusion}\label{sec:conclusion}

This paper has introduced \system, a database system that manages prompts as tuple-level \prompttype values. Unlike application-level prompting or LLM UDFs, \system makes prompts visible to the
database as typed, self-describing attributes with templates, bindings, model metadata, and task metadata. Users query generated evaluation views where prompt fields appear as ordinary SQL columns; internally, \system evaluates them through \texttt{EVAL}, applies database-guided rewrites, optionally selects a rewrite plan through \optimizer, and calls an LLM.

The central insight is that prompt execution has a database
optimization space. Constraints can improve validity, projection can reduce token cost, output minimization can improve parseability, and database-selected examples can improve task quality, especially for smaller language models. Treating these transformations as database rewrite rules allows \system to manage prompts systematically rather than relying on ad hoc application-level prompt engineering.

Future work includes extending the prompt context with foreign-key paths, database navigation, workload history, and provenance. We also intent to investigate learned cost and prompt quality models for enhancing database-guided prompt rewrite and optimization.


\bibliographystyle{ACM-Reference-Format}
\bibliography{references}

\end{document}